\documentclass[conference]{IEEEtran}
\IEEEoverridecommandlockouts
\usepackage{cite}
\usepackage{amsmath,amssymb,amsfonts}
\usepackage{subcaption}
\usepackage{graphicx}
\usepackage{caption}
\usepackage{multirow}
\usepackage[Procedure]{algorithm}
\usepackage[noend]{algpseudocode}
\usepackage{graphicx}
\usepackage{textcomp}
\usepackage{xcolor}
\def\BibTeX{{\rm B\kern-.05em{\sc i\kern-.025em b}\kern-.08em
    T\kern-.1667em\lower.7ex\hbox{E}\kern-.125emX}}

\definecolor{orange}{rgb}{1,0.5,0}

\usepackage{ulem}  
\usepackage{url}

\usepackage{booktabs}

\begin{document}

\title{Performance Comparison for Scientific Computations on the Edge via Relative Performance\\
\thanks{Financial support from the Deutsche Forschungsgemein- schaft (German Research Foundation) through grants GSC 111 and IRTG 2379 is gratefully acknowledged.}
}

\author{\IEEEauthorblockN{Aravind Sankaran}
\IEEEauthorblockA{\textit{IRTG - Modern Inverse Problems} \\
\textit{RWTH Aachen University}\\
Aachen, Germany \\
aravind.sankaran@rwth-aachen.de}
\and
\IEEEauthorblockN{Paolo Bientinesi}
\IEEEauthorblockA{\textit{Department of Computer Science} \\
\textit{Ume\r{a} Universitet}\\
Ume\r{a}, Sweden \\
pauldj@cs.umu.se}
}

\maketitle
\thispagestyle{plain}
\pagestyle{plain}

\begin{abstract}

In a typical Internet-of-Things setting that involves scientific applications, a target computation can be evaluated in many different ways depending on the split of computations among various devices. On the one hand, different implementations (or algorithms)---equivalent from a mathematical perspective---might exhibit significant difference in terms of performance.
On the other hand, some of the implementations are likely to show similar performance characteristics. In this paper, we focus on analysing the performance of a given set of algorithms by clustering them into performance classes. To this end, we use a measurement-based approach to evaluate and score algorithms based on pair-wise comparisons; we refer to this approach as ``Relative performance analysis".
Each comparison yields one of three outcomes: one algorithm can be ``better", ``worse", or ``equivalent" to another; those algorithms evaluating to have ``equivalent'' performance are merged into the same performance class. 
We show that our clustering methodology facilitates algorithm selection with respect to more than one metric; for instance, from the subset of equivalently fast algorithms, one could then select an algorithm that consumes the least energy on a certain device.

\end{abstract}


\section{Introduction}

With the rise of heterogeneous-computing, the evaluation of a mathematical expression can possibly be split among various devices such as CPU-GPU, CPU-Raspbian, CPU-Smartphone, etc. The optimization of scientific computations involves identifying the part of the code that can be executed on the edge device and the part that can be offloaded to an accelerator (such as a GPU) or a server; based on the split of the target computation among devices, there could be myriads of mathematically equivalent algorithms. The efficient computation of mathematical expressions is critical, especially for latency-sensitive applications; for instance, a faster real-time video analytics algorithm facilitates improved response time for intelligent vehicle applications~\cite{connectedvehicles}. However, when such algorithms are run on a resource-constrained hardware, it becomes important to ensure that the executions are also energy efficient. Therefore, in an Internet-of-Things (IoT) setting, the code is typically optimized with respect to more than one performance criteria (such as least execution time and energy).  

In this paper, a given set of mathematically equivalent algorithms $\mathcal{A}$ is clustered into performance classes $\mathcal{C}_i \subseteq A$ where $i \in \{1,2, \dots, k\}$ and $k$ is the number of performance classes\footnote{$k$ is determined dynamically and is not specified at the start.}. All the algorithms in a class $\mathcal{C}_i$ exhibit equivalent performance (based on one of the performance criteria), and the algorithms in $\mathcal{C}_i$ perform better than those in $\mathcal{C}_{i+1}$. The subsets $\mathcal{C}_i$'s are specific to a given computing architecture, operating system, and run time settings. Such a clustering of algorithms into performance classes facilitates the selection of an algorithm based on additional performance criteria; for instance, if the algorithms were clustered based on execution time, then from the subset of fast algorithms, one could choose the algorithm that performs at-most $X$ floating point operations on an energy-constrained edge device.

\begin{figure}[h!]
	\centering
		\begin{subfigure}[b]{0.45\textwidth}
		\includegraphics[width=0.85\linewidth]{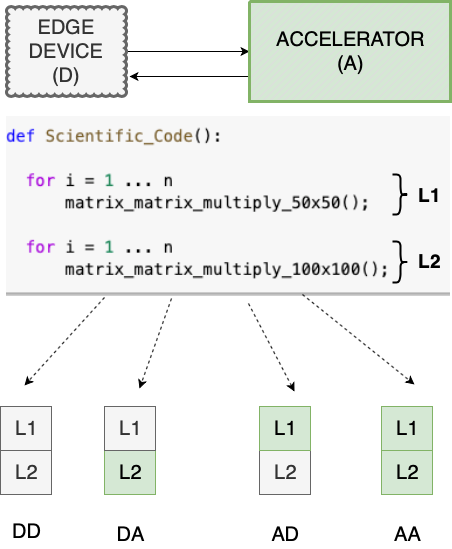}
		\caption{Different ways of splitting computations among devices}
		\label{fig:diff} 
	\end{subfigure}

	\begin{subfigure}[b]{0.45\textwidth}
		\includegraphics[width=0.9\linewidth]{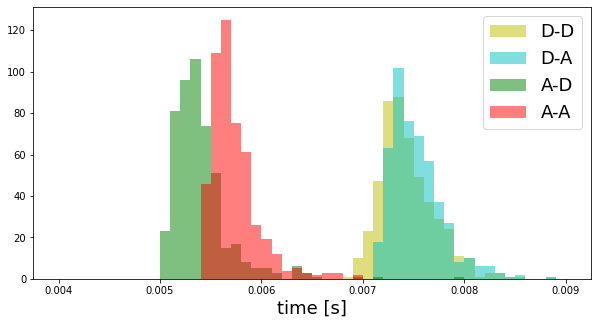}
		\caption{Distributions of execution times of the equivalent algorithms with computations split between CPU (D) and GPU (A)}
		\label{fig:eq}
	\end{subfigure}
	
	\caption{Splitting parts of a scientific code among devices.}
	\label{fig:1}
\end{figure}

The algorithms in $\mathcal{A}$ represent different, alternative ways of splitting a target computation among various device. In exact arithmetic, those algorithms would all return the same quantity. For instance, consider the ``scientific code'' in Figure \ref{fig:diff} consisting of two separate loops  L1 and L2  (assume that L2 cannot be executed before the completion of L1), each calling a certain function that performs matrix-matrix multiplication. Let us assume that the code is invoked from an edge device (D) and we have an option to offload a part of (or the whole) computation to an accelerator (A); if one restricts the granularity of offloading to an entire loop, then there are already four different ways to split the computations among the devices (shown in Figure \ref{fig:diff}). That is, either both L1 and L2 can be executed in the edge device (``DD'') or both offloaded to an accelerator (``AA''); or just one of the loops can be offloaded to an accelerator (``DA'' or ``AD'').

In order to score the algorithms in $\mathcal{A}$ and cluster them into performance classes, we use a measurement-based approach. Performance measurements such as execution time, are usually influenced by many factors, and that repeated measurements often result in different numbers~\cite{peise2014cache,hoefler2010characterizing,peise2012performance,rowanmemory}. As the variability in measurements increases (due to system noise), a single number (such as statistical mean, median or minimum) cannot reliably capture the performance of an algorithm; as a consequence, the result of algorithm comparison might not be consistent when the performance measurements are repeated.  Therefore, comparing the performance of any two algorithms should imply comparing two sets of measurements (or ``distributions''), and the result should yield one of three outcomes: ``better'', ``worse'' or ``equivalent''; in order for one algorithm to be better or worse than the other, there should be a significant difference in their distributions (for example, distributions ``AD'' and ``DD''  in Figure \ref{fig:eq}), otherwise, the two algorithms are said to be equivalent (for example, distributions ``DD'' and ``DA'' in Figure \ref{fig:eq}). 

We use the three-way pair-wise comparison to sort the algorithms in $\mathcal{A}$ according to the bubble-sort procedure~\cite{bubblesort}. In order to cluster the algorithms in $\mathcal{A}$ into performance classes, those algorithms evaluating to have equivalent performance in the pair-wise comparison are merged into the same class. The distributions of execution times of the four algorithms (implemented in Tensorflow 2.1~\cite{tensorflow}), with computations split between a CPU (Intel(R) Xeon(R) Platinum 8160, 1 core) and a GPU (Nvidia Pascal P100 SXM2), is shown in Figure \ref{fig:eq}. One could observe that the algorithm with just L1 offloaded to the GPU (``AD'') performs significantly better than the rest; although the L2 consists of a larger matrix-matrix multiplication (which in general should perform better in a GPU), for this particular example, the overhead caused by the larger data-movement between CPU and GPU is slightly more than the speed-up gain. In this paper, we do not make any assumption about the statistical nature of the distribution of performance measurements; thus, the approach extends naturally to any Device-Accelerator(s) combinations (such as CPU-Raspbian, Smartphone-GPU(s) etc.). The analysis can also be extended to any arbitrary mathematical function beyond matrix-matrix multiplication.

 Scientific codes consisting of a sequence of loops, each evaluating a certain mathematical expression, can be commonly encountered in many practical applications. We describe a couple of them:
\begin{enumerate}
	\item \textbf{Digital-Twin applications involving multi-scale modelling.} Evaluation of mathematical expressions in a loop is often encountered in simulation problems. Solving a  hierarchy of such problems (where results from one simulation are used to solve the next one) with varying computational volumes is known as multi-scale modelling. Some applications of multi-scale modelling include addressing challenges in highly contemporary issues such as climate, energy, biology, sociology, finance, etc.~\cite{hoekstra2014multiscale}.  In recent times, many real-time, latency-critical applications that involve multi-scale modelling are being realized because of the emergence of concepts such as Digital-Twin~\cite{nikula2020towards}. A digital representation that infers real-time updates of a physical object or a system, mostly by solving simulation problems, is known as a Digital-Twin. Real-time evaluation of complex simulations has become possible only in the recent times, and this is because of neural-network integration with multi-scale modelling~\cite{alber2019integrating}; the knowledge of physics is used to build such neural networks, and these methods are commonly known as Physics Informed Neural Networks~\cite{raissi2019physics}. For many such applications, scientific computations are performed on data that is collected from many devices (sensors, mostly constrained with computational capacity) and the split of computations among the computing resources has to be optimized to minimize response time. In~\cite{nikula2020towards}, the authors highlight about the need to address the computational  and hardware related challenges that arise in the practical adaptation of such applications. 
	
	\item \textbf{Hierarchical object-detection in images.}  A wide range of real-life applications ranging from autonomous driving to security in work environments have been realized with real-time object-detection~\cite{connectedvehicles}. The on-board processor in a hand-held device (such as a smartphone) or an autonomous drone, although constrained in computational capacity, can still be used to run low-fidelity object detectors (such as YOLO~\cite{huang2018yolo}) for quick identification of objects. However, higher fidelity object detectors (such as SSD~\cite{womg2018tiny}) can run simultaneously in the background and can be used to correct the low-fidelity detections (in case of error), but with a lag. This lag can be minimized by properly choosing the parts of the code that could be offloaded to an accelerator.   
\end{enumerate}

For our experiments, we consider a scientific code that solves a sequence of mathematical problems implemented in Tensorflow 2.1~\cite{tensorflow}.
Specifically, we consider a mathematical problem that consist of solving the Regularized Least Squares equation~\cite{golub1999tikhonov} in a loop. There are more than one such loops and each can be executed either on a CPU or a GPU\footnote{All the other device-accelerator settings  such as CPU-Smartphone, CPU-Raspbian etc., can be simulated by adding artificial delays and controlling the number of threads.}. We evaluate the performance of the algorithms by measuring the execution time, and clustering them into performance classes. We show that such a clustering can aid in selecting  the algorithm that minimizes the operating cost or the number of floating point operations performed (which minimizes energy) on a particular device. 


\textbf{\textit{Contributions:} } We use the methodology we developed in \cite{sankaran2021robust} to cluster a given set of equivalent algorithms $\mathcal{A}$. In contrast to our analysis in \cite{sankaran2021robust}, where we aimed to identify only the subset of ``fastest'' algorithms, we now focus on clusters of algorithms from all performance classes. One of the reasons for looking beyond just the fastest algorithms is that, in an edge computing environment, one might be interested in switching between algorithms from different performance classes to optimize energy in a device; for instance,  when energy consumption of a particular device reaches a certain threshold, one might be interested in switching to an algorithm that performs fewer floating point operations (FLOPs) on that device, and then switches back to the high-performance algorithm after a while. In this work, the execution times of the algorithms are explicitly measured and the resulting clusters are specific to the underlying architecture and run time settings; if the operating conditions are changed, the measurements have to be repeated. However, these clusters can be used as ground truth to train performance models that can automatically identify the algorithm of required performance without executing them\footnote{The modelling and prediction of relative performance is the objective of our future work, and it is out of the scope of this article.}. Moreover, such performance models for automatic algorithm selection can obtain better accuracy when trained with a particular loss function, known as Triplet loss~\cite{hermans2017defense}, where both positive (fast algorithm) and negative (worst algorithm) example are used to train the model; for such a training, the algorithms clustered into different performance classes would be required. 

\textit{\textbf{Organization}}: In Sec.~\ref{sec:rel}, we survey the state of art. In
Sec.~\ref{sec:met}, we describe the methodology
to cluster the equivalent algorithms and discuss some applications in Sec.~\ref{sec:exp}.
 Finally, in Sec.~\ref{sec:con}, we
draw conclusions and discuss the need for relative performance modelling.

\section{Related Works}
\label{sec:rel}
In the past decade, we have witnessed an increasing number of latency sensitive applications, such as intelligent vehicles\cite{connectedvehicles}, real-time video analytics\cite{videoanalytics}, augmented reality\cite{arvr} etc.,  that require costly scientific computations on resource-constrained hardware. For such applications, offloading all the computations to a cloud is not an option because of unacceptable communication latency between the cloud and end devices\cite{surveyMCC} \cite{towardsEdgeComputing}. On the other hand, increasing capabilities of mobile devices make them  more suitable for scientific computing loads\cite{smartPhonesForScientificComputing} \cite{raspberryEdgeComputing}. Consequently, there is a trend of computation offloading\cite{surveyOfComputationalOffloading2013} toward edge computing \cite{edgeComputing2016,towardsEdgeComputing,edgeComputing2015}, where significant gains are realized when computations are shifted towards the edge of the network (including the local device)\cite{EdgeComputingQuantifying}. For instance, a proof-of-concept platform that runs a face recognition application in \cite{facerecog} shows that the response time is reduced from 900 to 169 ms by moving computation from cloud to the edge. Depending on the distribution of workload among devices, myriads of implementations for the same computational problem can be devised, each having a significant impact on both latency and energy consumption. For instance,  Clone cloud in \cite{clonecloud} does on-demand partitioning of workload between mobile and the cloud, and their prototype could reduce 20x running time and energy for the tested application.

The increase of such decentralized applications in recent times can be attributed to containerization tools like Docker\cite{docker} that ease portability of code without compromising performance\cite{dockerForEdgeComputing} \cite{dockerhpc} and enable the use of same Linux environment across heterogeneous devices. Furthermore, recent high level programming languages and environments such as Julia\cite{julia}, Tensorflow\cite{tensorflow}, PyTorch\cite{pytorch} etc that abstract the calls to linear algebra libraries specially optimized for different hardware (such as raspberry Pi, GPU), allow developers to write code without worrying about the performance that can be achieved on the underlying architecture. Therefore, the same scientific code can achieve high performance on different devices, irrespective of the operating system, and the computations can easily be moved among devices just by modifying a flag value.  

The improved ease in developing high performance scientific code pushes academicians into new research directions that leads to realization of many real-time applications, which earlier seemed impractical. For instance,  traditional physics-based simulations are slow, and are used only to design, gain insights into system behaviour, or to analyse outputs and derive theories.  However, with the increasing adaptation of physics with fast neural network computations -- commonly referred by the umbrella term ``Physics Informed Neural Networks'' (PINNs)~\cite{raissi2019physics} -- significant speed ups in solving solving complex simulation problems have been reported off-late~\cite{kim2020fast,meng2020ppinn}. In \cite{willard2020integrating}, the authors survey and report a wide range of new technological applications with PINNs. In general, PINNs integrate data and physics-based approaches to create a virtual representation of a real system, commonly known as a Digital twin, which are then synchronized to realize many practical applications involving failure prediction and maintenance response on a factory floor, remote monitoring of autonomous vehicles etc~\cite{kritzinger2018digital,liu2020review}.  Most of these applications include the use of sensors through which data are collected and the computations can be distributed among many devices. In \cite{nikula2020towards}, the authors study the feasibility of practical adaptation of such technologies and mention about the need to address the computational and hardware related challenges that arise in the real-time environment. 

As a first step towards addressing the challenges, a framework to quantify and interpret performance of scientific codes in an edge computing setting is essential. While the previous works such as \cite{chen2018optimized,you2016energy,cicconetti2020architecture} optimize performance by finding out where to offload an entire computational task, we focus more on optimally splitting a single computational task among devices. Unlike \cite{qiu2019online, ren2018distributed,clonecloud}, where the focus of performance optimization is on a specific kind of computational task or operating environment, we attempt to arrive at a methodology that is more generic for scientific computations in general. Although our approach is measurement-based (the implementations have to be executed and timed), the results of our methodology will form a basis to derive automatic (execution-less) computation splitting strategies with reinforcement learning, or serve as a ground truth for performance prediction using supervised learning.

\section{Methodology}
\label{sec:met}


Given a set of $p$ equivalent algorithm $\mathbf{alg}_1,\dots \mathbf{alg}_p \in \mathcal{A}$, we want to form performance clusters $\mathcal{C}_i \subseteq \mathcal{A}$ where $i \in \{1,\dots,k\}$ and $k$ $ (\le p)$ is the number of clusters. $k$  is determined dynamically and need not be known at the start. The performance of the algorithms in cluster $\mathcal{C}_i$ is expected to be greater than those in $\mathcal{C}_{i+1}$. To this end, we first execute and measure the performance (say execution time) of every algorithm in $\mathcal{A}$ multiple times (say $N$ times). When the fluctuations in the performance measurements are large, it becomes difficult the summarize the performance into a single number.  Instead of summarizing the performance statistic (such as mean or minimum execution time) of all the $N$ measurements into one number, multiple statistics are evaluated and compared on data that is randomly sampled from the $N$ measurements; this approach is commonly known as ``bootstrapping''.
This allows us to gain more information from the set of measurements (also referred as distributions or histogram) of execution times, which are then used for performance comparisons~\cite{sankaran2021robust}. The distributions are compared pair-wise and merged into the same cluster if the comparison of two algorithms evaluates to be performance-equivalent. Two algorithms are said to be equivalent if their distributions of measurements significantly overlap. The ``bootstrapping'' strategy is used to quantify the overlap of distributions of two algorithms and classify one to be ``better'', ``equivalent'' or ``worse'' than the other; the procedure to implement this comparison strategy is described in Section IV of \cite{sankaran2021robust}.

\begin{algorithm}[t]
	\caption{SortAlgs $(\mathcal{A})$ }
	\label{alg:sort}
	\hspace*{\algorithmicindent} \textbf{Input: } $ \mathbf{alg}_1,\mathbf{alg}_2,\dots,\mathbf{alg}_p \in \mathcal{A}$ \\
	\hspace*{\algorithmicindent} \textbf{Output: } $ \langle (\mathbf{alg}_{s[1]},\textbf{rank}_1), \dots, (\mathbf{alg}_{s[p]},\textbf{rank}_p) \rangle $
	\begin{algorithmic}[1] 
		\For{i = 1, $\dots$, p}
		\State  Initialize $\mathbf{rank}_i \leftarrow i$ \Comment{Initialize Alg rank}
		\State  Initialize $\mathbf{s}_i \leftarrow i$ \Comment{Initialize Alg Indices}
		\EndFor
		\State Initialize $\mathcal{S} \leftarrow \langle (\mathbf{alg}_{s[1]},\textbf{rank}_1), \dots, (\mathbf{alg}_{s[p]},\textbf{rank}_p) \rangle $
		\For{i = 1, $\dots$, p}
		\For{j = 0, $\dots$, p-i-1}
		\State Compare $\mathbf{alg}_{s[j]}$ and $\mathbf{alg}_{s[j+1]}$
		\State $\mathcal{S} \leftarrow $ UpdateAlgIndices($\mathcal{S}, s[j]$)
		\State $\mathcal{S} \leftarrow $ UpdateAlgRanks($\mathcal{S}, s[j]$)
		\EndFor
		\EndFor
		\State return $\langle (\mathbf{alg}_{s[1]},\textbf{rank}_1), \dots, (\mathbf{alg}_{s[p]},\textbf{rank}_p) \rangle$
	\end{algorithmic}
\end{algorithm}

\begin{algorithm}[h!]
	\caption{UpdateAlgIndices $(\mathcal{S}, i)$ }
	\label{alg:upid}
	\hspace*{\algorithmicindent} \textbf{Input: } $ \langle (\mathbf{alg}_1, \textbf{rank}_1)\dots,(\mathbf{alg}_p,\textbf{rank}_p) \rangle\in \mathcal{S}$ \\
	\hspace*{\algorithmicindent} \hspace*{\algorithmicindent} $\quad i\in \{1,\dots,p\}$ \\
	\hspace*{\algorithmicindent} \textbf{Output: } Updated sequence set $\mathcal{S}$
	\begin{algorithmic}[1] 
		\If{$\mathbf{alg}_i < \mathbf{alg}_{i+1}$}
		\State Swap positions of $\mathbf{alg}_i$ and $\mathbf{alg}_{i+1}$
		\EndIf
		\State return Updated sequence set $\mathcal{S}$
	\end{algorithmic}
\end{algorithm}

\begin{algorithm}[h!]
	\caption{UpdateAlgRanks $(\mathcal{S}, i)$ }
	\label{alg:uprnk}
	\hspace*{\algorithmicindent} \textbf{Input: } $ \langle (\mathbf{alg}_1, \textbf{rank}_1)\dots,(\mathbf{alg}_p,\textbf{rank}_p) \rangle\in \mathcal{S}$ \\
	\hspace*{\algorithmicindent} \hspace*{\algorithmicindent} $\quad i\in \{1,\dots,p\}$ \\
	\hspace*{\algorithmicindent} \textbf{Output: } Updated sequence set $\mathcal{S}$
	\begin{algorithmic}[1] 
		\If{$\mathbf{alg}_i \sim \mathbf{alg}_{i+1}$ and \textbf{rank}$_i$ $\ne$ \textbf{rank}$_{i+1}$}
		\State \textbf{rank}$_{i+1}$, $\dots$, \textbf{rank}$_p$ decreased by 1
		\EndIf
		
		\If{$\mathbf{alg}_i > \mathbf{alg}_{i+1}$ and \textbf{rank}$_i$ $\ne$ \textbf{rank}$_{i+1}$}
		\If{\textbf{rank}$_i$ $=$ \textbf{rank}$_{i-1}$}
		\State \textbf{rank}$_{i+1}$, $\dots$, \textbf{rank}$_p$ decreased by 1.
		\EndIf
		\EndIf
		
		\If{$\mathbf{alg}_i > \mathbf{alg}_{i+1}$ and \textbf{rank}$_i$ $=$ \textbf{rank}$_{i+1}$}
		\If{\textbf{rank}$_i$ $\ne$ \textbf{rank}$_{i-1}$}
		\State \textbf{rank}$_{i+1}$, $\dots$, \textbf{rank}$_p$ increased by 1.
		\EndIf
		\EndIf
		\State return Updated sequence set $\mathcal{S}$
	\end{algorithmic}
\end{algorithm}

\begin{figure}
	\includegraphics[width=0.47\textwidth]{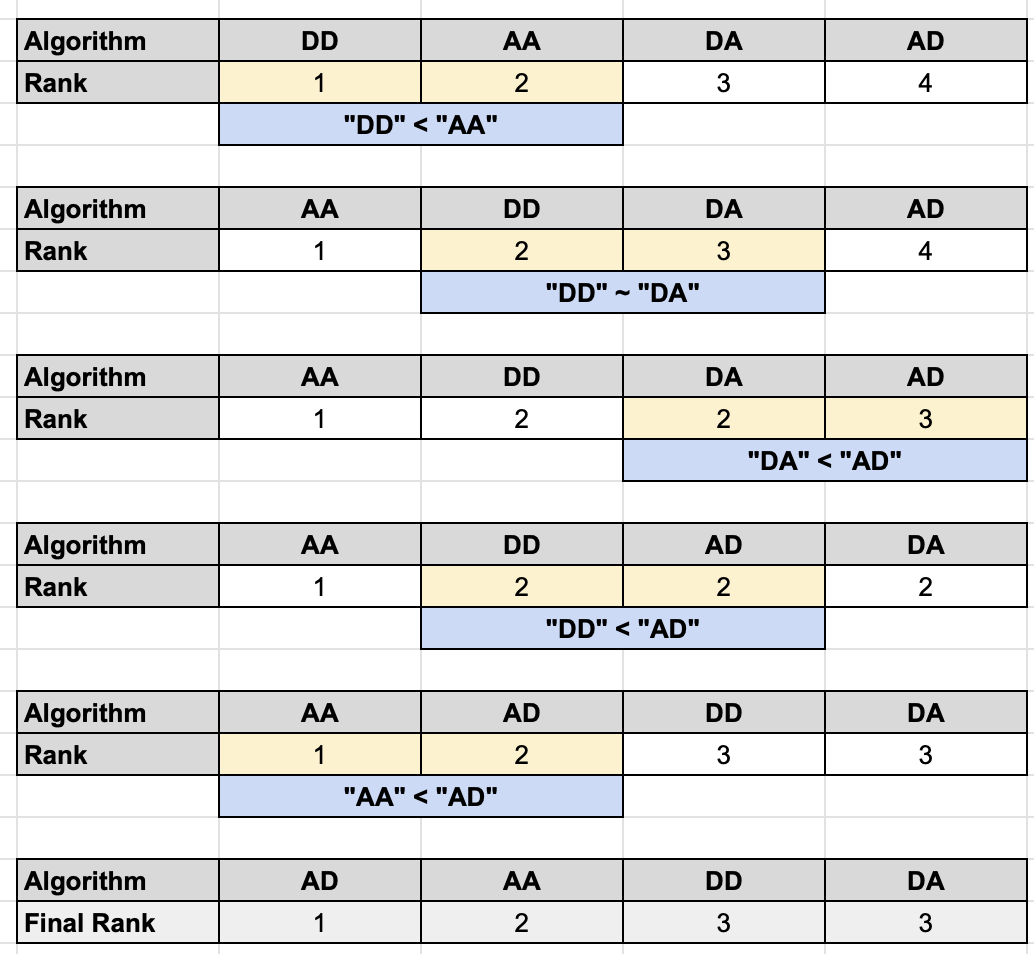}
	\caption{Bubble Sort with three-way comparison}
	\label{fig:sort}
\end{figure}

 We use the three-way comparison function to sort the set of algorithms in $\mathcal{A}$;  that is, we do a sort in which the comparison is not just a binary relation (better or worse), but also admits the equivalence of two algorithms.  This sorting procedure\footnote{In this work, we focus only on the update in each iteration of the sort and the sorting procedure as a whole is not optimized for performance. } is summarized in Procedure \ref{alg:sort}. The outcome of Procedure \ref{alg:sort} is a sequence set $\mathcal{S}$ consisting of tuples $\langle (\mathbf{alg}_{s[1]},\textbf{rank}_1), \dots, (\mathbf{alg}_{s[p]},\textbf{rank}_p) \rangle$ where $s[j]$ is the index of the $j^{th}$ algorithm in the sorted sequence and \textbf{rank}$_j \in \{1,\dots,k\}$ is the cluster assigned to $\mathbf{alg_j}$. In the beginning, the sequence set $\mathcal{S}$ is randomly initialized (see line 1-4 of Procedure \ref{alg:sort}) and the indices, and  the ranks of the algorithms are updated in every iteration of the sort (or bubble-sort) procedure. For instance, let us consider the illustration of clustering the four equivalent implementations of the scientific code $\mathbf{alg_{DD}},\mathbf{alg_{AA}},\mathbf{alg_{DA}},\mathbf{alg_{AD}} \in \mathcal{A}$ in Figure \ref{fig:diff}. The distributions of time measurements obtained by $N$=500 measurements of each algorithm are shown in Figure \ref{fig:eq}. Let the initial sequence set $\mathcal{S}$ be $\langle (\mathbf{alg_{s[1]=DD}},1), (\mathbf{alg_{s[2]=AA}},2), (\mathbf{alg_{s[3]=DA}},3),$  $(\mathbf{alg_{s[4]=AD}},4) \rangle$. The intermediate steps of the bubble sort procedure for this illustration is shown in Figure \ref{fig:sort}. In every iteration of the sort procedure, adjacent pairs of algorithms $\mathbf{alg}_{s[j]}$ and $\mathbf{alg}_{s[j+1]}$, starting from the right-most algorithm in $\mathcal{S}$, are compared using the bootstrap strategy from \cite{sankaran2021robust} and the following update rules are applied:
\begin{enumerate}
	\item \textbf{Update of Algorithm Indices: } An algorithm occurring earlier in the sorted sequence $\mathcal{S}$ should perform at least as good as those occurring later from it in the sequence. If an algorithm $\mathbf{alg}_{s[j]}$ performs worse ($<$) than its successor  $\mathbf{alg}_{s[j+1]}$, then the indices (or position) of both the algorithms in the sequence are swapped. Otherwise, the positions are left to remain the same. In the first comparison of our example, $\mathbf{alg_{s[1] = DD}}$ and $\mathbf{alg_{s[2] = AA}}$ are compared and $\mathbf{alg_{DD}}$ performs worse than its successor; hence, the two algorithms swap positions (now $\mathbf{alg_{s[1] = AA}}$ and $\mathbf{alg_{s[2] = DD}}$).
	\item \textbf{Update of algorithm ranks : }
	\begin{enumerate}
		\item \textbf{If the algorithms have not been swapped: } The algorithms are not swapped if  $\mathbf{alg}_{s[j]}$ evaluates to be better ($>$) or equivalent ($\sim$) to  $\mathbf{alg}_{s[j+1]}$. If the comparison is ``equivalent'', $\mathbf{alg}_{s[j]}$ and $\mathbf{alg}_{s[j+1]}$ should be assigned with the same rank; if the ranks are different, they are merged by decreasing the ranks of $\mathbf{alg}_{s[j+1]}, \dots \mathbf{alg}_{s[p]}$ by 1. If the comparison is ``better'', the ranks are not updated. In the second comparison of our illustration, the performance of $\mathbf{alg_{DD}}$ and $\mathbf{alg_{DA}}$ results to be ``equivalent'', but both have different ranks. Hence, the ranks of algorithms occurring after  $\mathbf{alg_{DD}}$ in the sequence are decreased by 1; $\mathbf{alg_{DD}}$ and $\mathbf{alg_{DA}}$ now share rank 2 and the rank of $\mathbf{alg_{AD}}$ is corrected to 3.   
		\item \textbf{If the algorithms have been swapped: } The indices of the algorithms are swapped only if $\mathbf{alg}_{s[j]}$ performs worse ($<$) than its successor $\mathbf{alg}_{s[j+1]}$. After swapping the positions of $\mathbf{alg}_{s[j]}$ and $\mathbf{alg}_{s[j+1]}$, 
		if ``$\mathbf{alg}_{s[j]}$ has the same rank as its predecessor $\mathbf{alg}_{s[j-1]}$, but different rank from its successor $\mathbf{alg}_{s[j+1]}$'', then the ranks of $\mathbf{alg}_{s[j+1]},\dots \mathbf{alg}_{s[p]}$ are decreased by 1.  However, if ``$\mathbf{alg}_{s[j]}$ shares the same rank with its successor $\mathbf{alg}_{s[j+1]}$, but has a different rank from its predecessor $\mathbf{alg}_{s[j-1]}$'', then the ranks of $\mathbf{alg}_{s[j+1]},\dots \mathbf{alg}_{s[p]}$ are increased by 1. For instance, in the third comparison in Figure \ref{fig:sort}, $\mathbf{alg_{s[3]=DA}}$ performs worse than $\mathbf{alg_{s[4]=AD}}$ and they swap positions -- now $\mathbf{alg_{s[3]=AD}}$ and $\mathbf{alg_{s[4]=DA}}$ -- and $\mathbf{alg_{s[3]=AD}}$ has the same rank as its predecessor $\mathbf{alg_{s[2]=DD}}$ (rank 2) but different from its successor $\mathbf{alg_{s[4]=DA}}$ (rank 3); hence the rank of  $\mathbf{alg_{s[4]=DA}}$ is decreased by 1, so that all the three algorithms $\mathbf{alg_{DD}}, \mathbf{alg_{AD}}, \mathbf{alg_{DA}}$ now share the same rank. In step 4 of our illustration, the algorithms with the same rank $\mathbf{alg_{s[2]=DD}}$ and $\mathbf{alg_{s[3]=AD}}$ swap positions, but after the swap $\mathbf{alg_{s[2]=AD}}$ has a different rank from its predecessor. Since $\mathbf{alg_{s[2]=AD}}$ reached the top of its performance class -- having defeated all the other algorithms with the same rank -- $\mathbf{alg_{s[2]=AD}}$ should now be assigned with a better rank than the other algorithms in its class. To this end,  $\mathbf{alg_{s[2]=AD}}$ stays at rank 2, but its successors $\mathbf{alg_{s[3]=DD}}$ and $\mathbf{alg_{s[4]=DA}}$ are pushed to rank 3.
	\end{enumerate} 
	
\end{enumerate}

Following the steps in Procedure \ref{alg:sort}, the final sequence set for our illustration results as $\langle (\mathbf{alg_{s[1]=AD}},1), (\mathbf{alg_{s[2]=AA}},2), (\mathbf{alg_{s[3]=DD}},3),$  $(\mathbf{alg_{s[4]=DA}},3) \rangle$, where the algorithms are clustered into three performance classes.


\paragraph*{\textbf{Computing the relative scores}}The clustering procedure discussed above is not deterministic, especially when the fluctuations in the performance measurements are large and the distributions are partially overlapping (for example, distributions ``AD'' and ``AA'' in Figure \ref{fig:eq}). Such overlaps become more evident when the number of measurements $N$ is small. For $N=30$, $\mathbf{alg_{AD}}$ is just at the threshold of being better than $\mathbf{alg_{AA}}$; hence, in the last step of  our sorting procedure, the result of comparing these two algorithms would switch between $\mathbf{alg_{AA}} < \mathbf{alg_{AD}}$ and $\mathbf{alg_{AA}} \sim \mathbf{alg_{AD}}$, and this can change the final rank of the algorithms. As $\mathbf{alg_{AA}}$ results to be equivalent to $\mathbf{alg_{AD}}$ once in every three comparisons,  approximately thirty percent of the times $\mathbf{alg_{AA}}$ is assigned to rank 1 cluster (and to the rank 2 cluster rest of the time). Therefore, we compute relative scores for the algorithms separately for each cluster. To this end, Procedure \ref{alg:sort} is repeated $Rep$ times\footnote{We do not repeat the execution of the algorithms, but repeat the procedure over the same set of $N$ measurements.} after shuffling the set $\mathcal{A}$ before each clustering and if an algorithm $\mathbf{alg}_j$ is assigned to cluster $r$ (or with rank $r$) in at least one out of $Rep$ iterations, $\mathbf{alg}_j$ will receive a ``relative score'' of $w_j/Rep$ (with respect to cluster $r$), where $w_j (\le Rep)$ is the number of times $\mathbf{alg}_j$ obtains rank $r$ (see Procedure \ref{alg:f}). The relative score of an algorithm $\mathbf{alg}_j$ with respect to cluster $r$ represents the confidence of $\mathbf{alg}_j$ in being assigned to cluster $r$. For our illustration, we obtain the following relative score estimates for algorithms assigned to each of the cluster:
\begin{itemize}
	\item Rank 1 cluster $\mathcal{C}_1 : \\ \{(\mathbf{alg_{AD}}, 1.0), (\mathbf{alg_{AA}}, 0.3)\}$
	\item Rank 2 cluster $\mathcal{C}_2 : \\ \{(\mathbf{alg_{AA}}, 0.7), (\mathbf{alg_{DD}}, 0.3), (\mathbf{alg_{DA}}, 0.3)\}$
	\item Rank 3 cluster $\mathcal{C}_3 : \\ \{ (\mathbf{alg_{DD}}, 0.7), (\mathbf{alg_{DA}}, 0.6)\}$
	\item Rank 4 cluster $\mathcal{C}_4 : \\ \{ (\mathbf{alg_{DA}}, 0.1)\}$
\end{itemize}
 The resulting clusters could be used as a ground truth to train performance models. If the assignment of algorithms to more than one cluster is not preferred for training performance models, then one could simply assign the algorithm to the cluster for which it obtained the maximum relative score and compute the final relative score by summing the scores from better ranks; for instance, $\mathbf{alg_{DA}}$ was assigned rank 3 in 60 percent of the iterations and rank 2 in 30 percent of the iterations, and hence $\mathbf{alg_{DA}}$ would be assigned with rank 3, for which it obtained the maximum relative score of 0.6, and its relative score from rank 2 would be cumulated, thereby resulting in a final relative score of 0.9.  Then, the final clustering would be $\mathcal{C}_1 : \{(\mathbf{alg_{AD}}, 1.0)\}$;  $\mathcal{C}_2 : \{(\mathbf{alg_{AA}}, 1.0)\}$;  $\mathcal{C}_3 : \{(\mathbf{alg_{DD}}, 1.0), (\mathbf{alg_{DA}}, 0.9)\}$.

\begin{algorithm}
	\caption{ GetCluster$_r$$(\mathcal{A}, Rep, R)$ }
	\label{alg:f}
	\hspace*{\algorithmicindent} \textbf{Input: } $ \mathbf{alg_1},\mathbf{alg_2} ,\dots, \mathbf{alg_p}\in \mathcal{A} \quad Reps,R \in \mathbb{Z}^{+}$ \\
	\hspace*{\algorithmicindent} \textbf{Output: } $ (\mathbf{a}_1,w_1), (\mathbf{a}_2, w_2), \dots, (\mathbf{a}_q,w_q) \in \mathcal{C}_r \times \mathbb{R}  $
	\begin{algorithmic}[1] 
		\State $\mathbf{a}\leftarrow [ \quad ]$ \Comment{Initialize empty lists}
		\State $\mathbf{a}_{\text{tmp}} \leftarrow [ \quad ]$ 
		\For{i = 1, $\dots$, $Rep$}
		\State Shuffle($\mathcal{A}$)
		\State SortAlgs$(\mathcal{A})$ 
		\State $\tilde{\mathbf{a}}_{\text{tmp}} \leftarrow$ select algorithms with rank $R$
		\State append $\tilde{\mathbf{a}}_{\text{tmp}}$ to the list $\mathbf{a}_{\text{tmp}} $
		\EndFor
		\State $\mathbf{a} \leftarrow $ select unique algorithms in $\mathbf{a}_{\text{tmp}}$ \Comment{$|\mathbf{a}| \le |\mathbf{a}_{\text{tmp}}|$}  
		\State $q = |\mathbf{a}|$
		\For{i = 1, $\dots$,$ q$ }
		\State $w_i \leftarrow$ number of occurrences of $\mathbf{a}_i$ in $\mathbf{a}$
		\State $w_i \leftarrow w_i/Rep$
		\EndFor
		\State return $ (\mathbf{a}_1,w_1), (\mathbf{a}_2, w_2), \dots, (\mathbf{a}_q,w_q) $ 
	\end{algorithmic}
\end{algorithm}

\section{Experiments}
\label{sec:exp}
The main motivation behind clustering of equivalent algorithms into performance classes is to aid in algorithm selection. The strategy to select the best algorithm, especially in an edge-computing environment, is not always based on just one performance metric (such as selecting the algorithm with the minimum execution time). For instance, if a computationally intensive algorithm is running on a resource constrained edge device such as a smart phone or a tablet, in addition to aiming for a fast execution of the code, it also becomes essential to monitor the resource usage from time to time so that the device does not consume more energy than the prescribed budget. One way to control the resource utilization on a device is by restricting the number of floating point operations (FLOPs) performed by the scientific code on that device. To this end, one can choose an alternate equivalent algorithm that restricts the FLOPs on that particular device by shipping parts of the computation to an accelerator. Now, the problem of algorithm selection arises when one has to decide on which parts of the computation should be exported to an accelerator so that the compromise on the overall execution time of the code is minimum. Recall that based on the split of computations, there can be many alternate algorithms with different performance characteristics. In this section, we demonstrate potential usages of our clustering methodology in facilitating algorithm selection. 

For our experiment, we consider a scientific code that calls a sequence of three ``$MathTasks$'' (\textbf{L1, L2, L3} in Procedure \ref{alg:sc}). The computational volume (number of FLOPs) of every $MathTask$ is different and every task computes a ``penalty'' that is used to solve the next task; hence, the three tasks cannot be executed concurrently. This kind of set-up is common in multi-scale modelling~\cite{hoekstra2014multiscale}. For our analysis, we consider a $MathTask$ that solves the Regularized Least Squares (RLS) equation in a loop (see Procedure \ref{alg:gls}) and the size of the matrices in the RLS equation determine the computational volume of the $MathTask$; higher size requires more FLOPs for computation. Each $MathTask$ can be computed either on a single core of Intel(R) Xeon(R) Platinum 8160 CPU (Edge device: D) or offloaded to Nvidia Pascal P100 SXM2 GPU (Accelerator: A). Although the GPU can perform more FLOPs per second, it has an extra cost due to data movement; therefore, offloading the entire code to the accelerator does not necessarily reduce the overall execution time. Depending on which part of the code -- \textbf{L1}, \textbf{L2} or \textbf{L3} -- is offloaded to the accelerator, there are totally 8 possible equivalent algorithms (see Table \ref{tab:rc}).  The execution time of every algorithm\footnote{ The code (implemented in Tensor Flow 2.1~\cite{tensorflow}) is available at ``https://github.com/HPAC/Relative-Performance''.} is measured 30 times and by applying the methodology described in Section~\ref{sec:met}, the distributions of measurements have been clustered into five performance classes. Table \ref{tab:rc} shows the relative scores of the algorithms in each cluster.  

\begin{algorithm}[ht]
	\caption{Scientific Code (\textbf{D1, D2, D3}) }
	\label{alg:sc}
	\hspace*{\algorithmicindent} \textbf{Input: } \textbf{D1, D2, D3 } $\in $ $\{$device (D) , accelerator (A)$\}$
	\begin{algorithmic}[1]
		\State penalty $\leftarrow 0$
		\State Run on \textbf{D1}:
		\State \hspace*{\algorithmicindent} penalty $\leftarrow$ $MathTask$ (50, penalty) \Comment{\textbf{L1}}
		\State Run on \textbf{D2}:
		\State \hspace*{\algorithmicindent} penalty $\leftarrow$ $MathTask$ (75, penalty) \Comment{\textbf{L2}}
		\State Run on \textbf{D3}:
		\State \hspace*{\algorithmicindent} penalty $\leftarrow$ $MathTask$ (300, penalty) \Comment{\textbf{L3}}
	\end{algorithmic}
\end{algorithm}
\begin{algorithm}[h!]
	\caption{ $MathTask$ (size, penalty) }
	\label{alg:gls}
	\hspace*{\algorithmicindent} \textbf{Input: } size, penalty $\in \mathbb{R}$
	\begin{algorithmic}[1] 
	\For{i=1, $\dots$, n} 
	\State Randomly generate $A$ \Comment{$A \in \mathbb{R}^{\text{size} \times \text{size}}$}
	\State Randomly generate $B$ \Comment{$B \in \mathbb{R}^{\text{size} \times \text{size}}$}
	\State 	$Z \leftarrow (A^{T}A + \text{penalty.}I)^{-1}A^{T}B$ 
	\State penalty $ \leftarrow ||AZ-B||^{2}$
	\EndFor
	\State return penality
	\end{algorithmic}
\end{algorithm}
\begin{table}[h!]
	\begin{center}
		\renewcommand{\arraystretch}{1.2}
		\begin{tabular}{@{}ccc@{}}
			\toprule
			Cluster & Algorithm  & Relative Score  \\
			\toprule
			 $\mathcal{C}_1$ & $\mathbf{alg_{DDA}}$ & 1.0 \\
						   & $\mathbf{alg_{DAA}}$ & 0.6 \\
			\midrule
			 $\mathcal{C}_2$ &   $\mathbf{alg_{DDD}}$ & 1.0 \\
			    &  $\mathbf{alg_{DAA}}$ & 0.4 \\
			  \midrule
			  $\mathcal{C}_3$ &  $\mathbf{alg_{ADA}}$ & 1.0 \\
			  &  $\mathbf{alg_{ADD}}$ & 1.0 \\
			  &  $\mathbf{alg_{DAD}}$ & 0.7 \\
			  \midrule
			  $\mathcal{C}_4$ &  $\mathbf{alg_{AAA}}$ & 1.0 \\
			  &  $\mathbf{alg_{DAD}}$ & 0.3 \\
			  \midrule
			  $\mathcal{C}_5$ &  $\mathbf{alg_{AAD}}$ & 1.0 \\
			\bottomrule
		\end{tabular}
		\caption{Clustering of algorithms}
		\label{tab:rc}
	\end{center}
\end{table}

Let us now assume that there is an operating cost involved in executing the code on the accelerator (A). In order to minimize such a cost, the best option would be to execute the whole code on the edge device (D). But executing the entire code on the edge device ($\mathbf{alg_{DDD}}$) is not the optimum choice in terms of performance; yet $\mathbf{alg_{DDD}}$ is not so bad, as it results in the second best performance class $\mathcal{C}_2$.
Procuring an accelerator and offloading \textbf{L3} to it would improve performance, as algorithm $\mathbf{alg_{DDA}}$ results in the best performance class $\mathcal{C}_1$.  Whether one should spend money  on an accelerator or not would depend on the margin of speed up and what the application as a whole can gain through that speed up. Therefore, the choice of algorithm is now based on a ``decision-model'' that is a trade-off between operating cost and speed. In our example,  for a small loop size of $n=10$ in Procedure \ref{alg:gls}, the mean execution time of $\mathbf{alg_{DDA}}$ is just 0.002s more than $\mathbf{alg_{DDD}}$ and the speed up is approximately 1.05. When $n$ becomes larger, the speed up increases. Thus, a decision-model can make a trade-off between $n$, relative scores and operating cost to aid in algorithm selection; the weight on the operating cost would depend on the importance of speed-up for the application. For latency critical applications such as those involving  response of the autonomous vehicle to external conditions, even a small improvement of 0.002s in the execution time can make a significant difference in the quality of the application; for instance, the safety of autonomous vehicles can improve if an object detector can evaluate more objects per second. 

Let us consider another application where it is ideal to run the whole code on the edge device ($\mathbf{alg_{DDD}}$); however, the device cannot persistently handle all the computations because of energy constraints. Therefore, in regular intervals, the amount of computations on the edge has to be reduced for a small period of time. In such a case, one can switch to $\mathbf{alg_{DAA}}$ among the algorithms in $\mathcal{C}_1$, as it offloads most of the computations to the accelerator, and then switch back to $\mathbf{alg_{DDD}}$ when the device cools down.

\section{Conclusion and Future Outlook}
\label{sec:con}

Efficient evaluation of scientific computations on the Edge is essential for latency-sensitive application such as intelligent vehicles~\cite{connectedvehicles}, augmented reality~\cite{arvr} etc. For a given scientific computation, we considered all the equivalent solution algorithms based on the split of computation among devices; these solution algorithms are mathematically equivalent to one another, but can exhibit significant difference in performance. We presented a measurement-based approach to cluster these algorithms into performance classes through pair-wise comparison of performance measurements. We showed that such a clustering can aid in algorithm selection based on additional criteria such as operating cost and energy.   

Although we considered a use-case where it is possible to execute and measure all the different combination of solution algorithms, this may not be an ideal solution for applications where there is an exponential number of alternative implementations. For instance, the linear algebra expression in the line 4 of Procedure \ref{alg:gls} can alone have many different equivalent algorithms, each having a different sequence of calls to optimized libraries such as BLAS and LAPACK~\cite{psarras2019linear}; typically these algorithms also show significant difference in performance, even without considering the split of computation among devices~\cite{sankaran2021robust} \cite{barthels2019linnea}.  Therefore, in case of exponential explosion of the search space, our methodology can still be applied on a subset of possible solutions and the resulting clusters with relative scores can be used as a ground truth to guide the search of algorithm via reinforcement learning. Thus, our methodology can be used to develop performance models that predict relative scores without having to execute all the algorithms. 

\bibliographystyle{IEEEtran}
\bibliography{paise21}

\begin{thebibliography}{10}
\providecommand{\url}[1]{#1}
\csname url@samestyle\endcsname
\providecommand{\newblock}{\relax}
\providecommand{\bibinfo}[2]{#2}
\providecommand{\BIBentrySTDinterwordspacing}{\spaceskip=0pt\relax}
\providecommand{\BIBentryALTinterwordstretchfactor}{4}
\providecommand{\BIBentryALTinterwordspacing}{\spaceskip=\fontdimen2\font plus
\BIBentryALTinterwordstretchfactor\fontdimen3\font minus
  \fontdimen4\font\relax}
\providecommand{\BIBforeignlanguage}[2]{{%
\expandafter\ifx\csname l@#1\endcsname\relax
\typeout{** WARNING: IEEEtran.bst: No hyphenation pattern has been}%
\typeout{** loaded for the language `#1'. Using the pattern for}%
\typeout{** the default language instead.}%
\else
\language=\csname l@#1\endcsname
\fi
#2}}
\providecommand{\BIBdecl}{\relax}
\BIBdecl

\bibitem{connectedvehicles}
D.~Grewe, M.~Wagner, M.~Arumaithurai, I.~Psaras, and D.~Kutscher,
  ``Information-centric mobile edge computing for connected vehicle
  environments: Challenges and research directions,'' in \emph{Proceedings of
  the Workshop on Mobile Edge Communications}, 2017, pp. 7--12.

\bibitem{peise2014cache}
E.~Peise and P.~Bientinesi, ``A study on the influence of caching: Sequences of
  dense linear algebra kernels,'' in \emph{International Conference on High
  Performance Computing for Computational Science}.\hskip 1em plus 0.5em minus
  0.4em\relax Springer, 2014, pp. 245--258.

\bibitem{hoefler2010characterizing}
T.~Hoefler, T.~Schneider, and A.~Lumsdaine, ``Characterizing the influence of
  system noise on large-scale applications by simulation,'' in \emph{SC'10:
  Proceedings of the 2010 ACM/IEEE International Conference for High
  Performance Computing, Networking, Storage and Analysis}.\hskip 1em plus
  0.5em minus 0.4em\relax IEEE, 2010, pp. 1--11.

\bibitem{peise2012performance}
E.~Peise and P.~Bientinesi, ``Performance modeling for dense linear algebra,''
  in \emph{2012 SC Companion: High Performance Computing, Networking Storage
  and Analysis}.\hskip 1em plus 0.5em minus 0.4em\relax IEEE, 2012, pp.
  406--416.

\bibitem{rowanmemory}
\BIBentryALTinterwordspacing
R.~Iakymchuk and P.~Bientinesi, ``Modeling performance through memory-stalls,''
  \emph{ACM SIGMETRICS Perform. Eval. Rev.}, vol.~40, no.~2, p. 86–91, Oct.
  2012. [Online]. Available: \url{https://doi.org/10.1145/2381056.2381076}
\BIBentrySTDinterwordspacing

\bibitem{bubblesort}
O.~Astrachan, ``Bubble sort: an archaeological algorithmic analysis,''
  \emph{ACM Sigcse Bulletin}, vol.~35, no.~1, pp. 1--5, 2003.

\bibitem{tensorflow}
M.~Abadi, P.~Barham, J.~Chen, Z.~Chen, A.~Davis, J.~Dean, M.~Devin,
  S.~Ghemawat, G.~Irving, M.~Isard \emph{et~al.}, ``Tensorflow: A system for
  large-scale machine learning,'' in \emph{12th $\{$USENIX$\}$ symposium on
  operating systems design and implementation ($\{$OSDI$\}$ 16)}, 2016, pp.
  265--283.

\bibitem{hoekstra2014multiscale}
A.~Hoekstra, B.~Chopard, and P.~Coveney, ``Multiscale modelling and simulation:
  a position paper,'' \emph{Philosophical Transactions of the Royal Society A:
  Mathematical, Physical and Engineering Sciences}, vol. 372, no. 2021, p.
  20130377, 2014.

\bibitem{nikula2020towards}
R.-P. Nikula, M.~Paavola, M.~Ruusunen, and J.~Keski-Rahkonen, ``Towards online
  adaptation of digital twins,'' \emph{Open Engineering}, vol.~10, no.~1, pp.
  776--783, 2020.

\bibitem{alber2019integrating}
M.~Alber, A.~B. Tepole, W.~R. Cannon, S.~De, S.~Dura-Bernal, K.~Garikipati,
  G.~Karniadakis, W.~W. Lytton, P.~Perdikaris, L.~Petzold \emph{et~al.},
  ``Integrating machine learning and multiscale modeling—perspectives,
  challenges, and opportunities in the biological, biomedical, and behavioral
  sciences,'' \emph{NPJ digital medicine}, vol.~2, no.~1, pp. 1--11, 2019.

\bibitem{raissi2019physics}
M.~Raissi, P.~Perdikaris, and G.~E. Karniadakis, ``Physics-informed neural
  networks: A deep learning framework for solving forward and inverse problems
  involving nonlinear partial differential equations,'' \emph{Journal of
  Computational Physics}, vol. 378, pp. 686--707, 2019.

\bibitem{huang2018yolo}
R.~Huang, J.~Pedoeem, and C.~Chen, ``Yolo-lite: a real-time object detection
  algorithm optimized for non-gpu computers,'' in \emph{2018 IEEE International
  Conference on Big Data (Big Data)}.\hskip 1em plus 0.5em minus 0.4em\relax
  IEEE, 2018, pp. 2503--2510.

\bibitem{womg2018tiny}
A.~Womg, M.~J. Shafiee, F.~Li, and B.~Chwyl, ``Tiny ssd: A tiny single-shot
  detection deep convolutional neural network for real-time embedded object
  detection,'' in \emph{2018 15th Conference on Computer and Robot Vision
  (CRV)}.\hskip 1em plus 0.5em minus 0.4em\relax IEEE, 2018, pp. 95--101.

\bibitem{golub1999tikhonov}
G.~H. Golub, P.~C. Hansen, and D.~P. O'Leary, ``Tikhonov regularization and
  total least squares,'' \emph{SIAM journal on matrix analysis and
  applications}, vol.~21, no.~1, pp. 185--194, 1999.

\bibitem{sankaran2021robust}
A.~Sankaran and P.~Bientinesi, ``Robust ranking of equivalent algorithms via
  relative performance,'' \emph{arXiv preprint arXiv:2010.07226}, 2021.

\bibitem{hermans2017defense}
A.~Hermans, L.~Beyer, and B.~Leibe, ``In defense of the triplet loss for person
  re-identification,'' \emph{arXiv preprint arXiv:1703.07737}, 2017.

\bibitem{videoanalytics}
G.~Ananthanarayanan, P.~Bahl, P.~Bod{\'\i}k, K.~Chintalapudi, M.~Philipose,
  L.~Ravindranath, and S.~Sinha, ``Real-time video analytics: The killer app
  for edge computing,'' \emph{computer}, vol.~50, no.~10, pp. 58--67, 2017.

\bibitem{arvr}
R.-S. Schmoll, S.~Pandi, P.~J. Braun, and F.~H. Fitzek, ``Demonstration of
  vr/ar offloading to mobile edge cloud for low latency 5g gaming
  application,'' in \emph{2018 15th IEEE Annual Consumer Communications \&
  Networking Conference (CCNC)}.\hskip 1em plus 0.5em minus 0.4em\relax IEEE,
  2018, pp. 1--3.

\bibitem{surveyMCC}
H.~T. Dinh, C.~Lee, D.~Niyato, and P.~Wang, ``A survey of mobile cloud
  computing: architecture, applications, and approaches,'' \emph{Wireless
  communications and mobile computing}, vol.~13, no.~18, pp. 1587--1611, 2013.

\bibitem{towardsEdgeComputing}
L.~{Lin}, X.~{Liao}, H.~{Jin}, and P.~{Li}, ``Computation offloading toward
  edge computing,'' \emph{Proceedings of the IEEE}, vol. 107, no.~8, pp.
  1584--1607, 2019.

\bibitem{smartPhonesForScientificComputing}
J.~M. Rodr{\'i}guez, C.~Mateos, and A.~Zunino, ``Are smartphones really useful
  for scientific computing?'' in \emph{Advances in New Technologies,
  Interactive Interfaces and Communicability}, F.~Cipolla-Ficarra, K.~Veltman,
  D.~Verber, M.~Cipolla-Ficarra, and F.~Kamm{\"u}ller, Eds.\hskip 1em plus
  0.5em minus 0.4em\relax Berlin, Heidelberg: Springer Berlin Heidelberg, 2012,
  pp. 38--47.

\bibitem{raspberryEdgeComputing}
M.~Maksimovi{\'c}, V.~Vujovi{\'c}, N.~Davidovi{\'c}, V.~Milo{\v{s}}evi{\'c},
  and B.~Peri{\v{s}}i{\'c}, ``Raspberry pi as internet of things hardware:
  performances and constraints,'' \emph{design issues}, vol.~3, no.~8, 2014.

\bibitem{surveyOfComputationalOffloading2013}
K.~Kumar, J.~Liu, Y.-H. Lu, and B.~Bhargava, ``A survey of computation
  offloading for mobile systems,'' \emph{Mobile networks and Applications},
  vol.~18, no.~1, pp. 129--140, 2013.

\bibitem{edgeComputing2016}
W.~Shi, J.~Cao, Q.~Zhang, Y.~Li, and L.~Xu, ``Edge computing: Vision and
  challenges,'' \emph{IEEE internet of things journal}, vol.~3, no.~5, pp.
  637--646, 2016.

\bibitem{edgeComputing2015}
P.~Garcia~Lopez, A.~Montresor, D.~Epema, A.~Datta, T.~Higashino, A.~Iamnitchi,
  M.~Barcellos, P.~Felber, and E.~Riviere, ``Edge-centric computing: Vision and
  challenges,'' 2015.

\bibitem{EdgeComputingQuantifying}
W.~Hu, Y.~Gao, K.~Ha, J.~Wang, B.~Amos, Z.~Chen, P.~Pillai, and
  M.~Satyanarayanan, ``Quantifying the impact of edge computing on mobile
  applications,'' in \emph{Proceedings of the 7th ACM SIGOPS Asia-Pacific
  Workshop on Systems}, 2016, pp. 1--8.

\bibitem{facerecog}
S.~Yi, Z.~Hao, Z.~Qin, and Q.~Li, ``Fog computing: Platform and applications,''
  in \emph{2015 Third IEEE Workshop on Hot Topics in Web Systems and
  Technologies (HotWeb)}.\hskip 1em plus 0.5em minus 0.4em\relax IEEE, 2015,
  pp. 73--78.

\bibitem{clonecloud}
\BIBentryALTinterwordspacing
B.-G. Chun, S.~Ihm, P.~Maniatis, M.~Naik, and A.~Patti, ``Clonecloud: Elastic
  execution between mobile device and cloud,'' in \emph{Proceedings of the
  Sixth Conference on Computer Systems}, ser. EuroSys '11.\hskip 1em plus 0.5em
  minus 0.4em\relax New York, NY, USA: Association for Computing Machinery,
  2011, p. 301–314. [Online]. Available:
  \url{https://doi.org/10.1145/1966445.1966473}
\BIBentrySTDinterwordspacing

\bibitem{docker}
D.~Merkel, ``Docker: lightweight linux containers for consistent development
  and deployment,'' \emph{Linux journal}, vol. 2014, no. 239, p.~2, 2014.

\bibitem{dockerForEdgeComputing}
B.~I. {Ismail}, E.~{Mostajeran Goortani}, M.~B. {Ab Karim}, W.~{Ming Tat},
  S.~{Setapa}, J.~Y. {Luke}, and O.~{Hong Hoe}, ``Evaluation of docker as edge
  computing platform,'' in \emph{2015 IEEE Conference on Open Systems (ICOS)},
  2015, pp. 130--135.

\bibitem{dockerhpc}
A.~Azab, ``Enabling docker containers for high-performance and many-task
  computing,'' in \emph{2017 ieee international conference on cloud engineering
  (ic2e)}.\hskip 1em plus 0.5em minus 0.4em\relax IEEE, 2017, pp. 279--285.

\bibitem{julia}
J.~Bezanson, A.~Edelman, S.~Karpinski, and V.~B. Shah, ``Julia: A fresh
  approach to numerical computing,'' \emph{SIAM review}, vol.~59, no.~1, pp.
  65--98, 2017.

\bibitem{pytorch}
A.~Paszke, S.~Gross, F.~Massa, A.~Lerer, J.~Bradbury, G.~Chanan, T.~Killeen,
  Z.~Lin, N.~Gimelshein, L.~Antiga \emph{et~al.}, ``Pytorch: An imperative
  style, high-performance deep learning library,'' in \emph{Advances in neural
  information processing systems}, 2019, pp. 8026--8037.

\bibitem{kim2020fast}
Y.~Kim, Y.~Choi, D.~Widemann, and T.~Zohdi, ``A fast and accurate
  physics-informed neural network reduced order model with shallow masked
  autoencoder,'' \emph{arXiv preprint arXiv:2009.11990}, 2020.

\bibitem{meng2020ppinn}
X.~Meng, Z.~Li, D.~Zhang, and G.~E. Karniadakis, ``Ppinn: Parareal
  physics-informed neural network for time-dependent pdes,'' \emph{Computer
  Methods in Applied Mechanics and Engineering}, vol. 370, p. 113250, 2020.

\bibitem{willard2020integrating}
J.~Willard, X.~Jia, S.~Xu, M.~Steinbach, and V.~Kumar, ``Integrating
  physics-based modeling with machine learning: A survey,'' \emph{arXiv
  preprint arXiv:2003.04919}, 2020.

\bibitem{kritzinger2018digital}
W.~Kritzinger, M.~Karner, G.~Traar, J.~Henjes, and W.~Sihn, ``Digital twin in
  manufacturing: A categorical literature review and classification,''
  \emph{IFAC-PapersOnLine}, vol.~51, no.~11, pp. 1016--1022, 2018.

\bibitem{liu2020review}
M.~Liu, S.~Fang, H.~Dong, and C.~Xu, ``Review of digital twin about concepts,
  technologies, and industrial applications,'' \emph{Journal of Manufacturing
  Systems}, 2020.

\bibitem{chen2018optimized}
X.~Chen, H.~Zhang, C.~Wu, S.~Mao, Y.~Ji, and M.~Bennis, ``Optimized computation
  offloading performance in virtual edge computing systems via deep
  reinforcement learning,'' \emph{IEEE Internet of Things Journal}, vol.~6,
  no.~3, pp. 4005--4018, 2018.

\bibitem{you2016energy}
C.~You, K.~Huang, H.~Chae, and B.-H. Kim, ``Energy-efficient resource
  allocation for mobile-edge computation offloading,'' \emph{IEEE Transactions
  on Wireless Communications}, vol.~16, no.~3, pp. 1397--1411, 2016.

\bibitem{cicconetti2020architecture}
C.~Cicconetti, M.~Conti, and A.~Passarella, ``Architecture and performance
  evaluation of distributed computation offloading in edge computing,''
  \emph{Simulation Modelling Practice and Theory}, vol. 101, p. 102007, 2020.

\bibitem{qiu2019online}
X.~Qiu, L.~Liu, W.~Chen, Z.~Hong, and Z.~Zheng, ``Online deep reinforcement
  learning for computation offloading in blockchain-empowered mobile edge
  computing,'' \emph{IEEE Transactions on Vehicular Technology}, vol.~68,
  no.~8, pp. 8050--8062, 2019.

\bibitem{ren2018distributed}
J.~Ren, Y.~Guo, D.~Zhang, Q.~Liu, and Y.~Zhang, ``Distributed and efficient
  object detection in edge computing: Challenges and solutions,'' \emph{IEEE
  Network}, vol.~32, no.~6, pp. 137--143, 2018.

\bibitem{psarras2019linear}
C.~Psarras, H.~Barthels, and P.~Bientinesi, ``The linear algebra mapping
  problem,'' \emph{arXiv preprint arXiv:1911.09421}, 2019.

\bibitem{barthels2019linnea}
H.~Barthels, C.~Psarras, and P.~Bientinesi, ``Linnea: Automatic generation of
  efficient linear algebra programs,'' \emph{arXiv preprint arXiv:1912.12924},
  2019.

\end{thebibliography}


\end{document}